\documentclass[english]{gretsi}

\usepackage[utf8]{inputenc}

\usepackage[T1]{fontenc}

\usepackage{amsmath, amssymb, amsfonts}

\usepackage[font=footnotesize,skip=0pt]{subcaption}

\usepackage{enumitem}

\usepackage{enumitem}
\usepackage{setspace}
\usepackage{xcolor}
\usepackage{comment}

\newcommand{\Reseau}{\mathcal{R}}
\newcommand{\Rcible}{\mathcal{R}^{\text{target}}}
\newcommand{\Rfant}{\mathcal{R}^{\text{shadow}}}
\newcommand{\Dentr}{\mathcal{D}_{\text{train}}}
\newcommand{\Dtest}{\mathcal{D}_{\text{test}}}
\newcommand{\mD}{\mathcal{D}}

\newcommand\matseq[2]{\mathbf{#1}^{(#2)}}

\newcommand\extrafootertext[1]{%
    \bgroup
    \renewcommand\thefootnote{\fnsymbol{footnote}}%
    \renewcommand\thempfootnote{\fnsymbol{mpfootnote}}%
    \footnotetext[0]{#1}%
    \egroup
}
\begin{document}


\titre{Can sparsity improve the privacy of neural networks?
}

\auteurs{
  \auteur{Antoine}{Gonon*}{antoine.gonon+gretsi@ens-lyon.fr}{1}
  \auteur{Léon}{Zheng*}{leon.zheng+gretsi@ens-lyon.fr}{1, 2}
  \auteur{Clément}{Lalanne}{clement.lalanne+gretsi@ens-lyon.fr}{1}
  \auteur{Quoc-Tung}{le}{quoc-tung.le+gretsi@ens-lyon.fr}{1}
  \auteur{Guillaume}{Lauga\dag}{guillaume.lauga+gretsi@ens-lyon.fr}{1}
  \auteur{Can}{Pouliquen\dag}{can.pouliquen+gretsi@ens-lyon.fr}{1}
}

\affils{
  \affil{1}{Univ Lyon, EnsL, UCBL, CNRS, Inria,  LIP, F-69342, LYON Cedex 07, France
  }
    \affil{2}{valeo.ai, Paris, France
  }
}

\abstract{Sparse neural networks are mainly motivated by ressource efficiency since they use fewer parameters than their dense counterparts but still reach comparable accuracies. This article empirically investigates whether sparsity could also improve the privacy of the data used to train the networks. The experiments show {\em positive correlations between the sparsity of the model, its privacy, and its classification error}. Simply comparing the privacy of two models with different sparsity levels can yield misleading conclusions on the role of sparsity, because of the additional correlation with the classification error. From this perspective, some caveats are raised about previous works that investigate sparsity and privacy.}

\maketitle


\section{Introduction}
\extrafootertext{*,\dag: Equal contributions. Work funded in part by the following projects: AllegroAssai ANR-19-CHIA-0009, NuSCAP ANR-20-CE48-0014, SeqALO ANR-20-CHIA-0020-01, MOMIGS of GdR ISIS and
CIFRE N°2020/1643. The authors thank the Blaise Pascal center for the computational means. It uses the SIDUS \cite{quemener2013} solution developed by Emmanuel Quemener.}

Deep neural networks are state-of-the-art for many learning problems. In practice, it is possible to tune the parameters of a given network in order to perfectly interpolate the available data \cite{Zhang21UnderstandingRequiresRethinking}. This overfitting regime is of practical interest since good performances can be obtained this way \cite{Belkin19DoubleDescent}. However, it comes with an increased risk in terms of privacy \cite{SurveyRigaki20}, since the network memorizes information about training data, up to the point of interpolating them. Among these information, some might be confidential. This raises the question of what information can be inferred given a black-box access to the model.

To detect an overfitting situation, an indicator is given by the ratio of the number of parameters by the number of data points available: the more parameters there are, the more the model can interpolate the data. In order to hinder the capacity of the model to overfit, and thus to store confidential information, this work studies the role of the number of nonzero parameters used. \emph{Can we find a good trade-off between model accuracy and privacy by tuning the sparsity (number of nonzero parameters) of neural networks?}

Attacks such as "Membership Inference Attack" (MIA) can infer the membership of a data point to the training set \cite{shokri2017membership}, using only a black-box access to the targeted model. This can be problematic in case of sensitive data (medical data, etc.). Given a network, how could one reduce the risk of such attacks, while preserving its performances as much as possible?

Numerous procedures have been proposed to defend against MIAs \cite{MIAsurvey2022}. In this work, the studied approach consists in decreasing the number of nonzero parameters used by the network in order to reduce its memorization capacity, while preserving as much as possible its accuracy. Decreasing the number of parameters can either be done by considering smaller (dense) networks as in \cite{tan2023blessing}, or by considering sparse subnetworks, as done here\footnote{One can empirically find sparse subnetworks that are more accurate than small dense networks: $\textrm{sparse subnetwork }\simeq \textrm{dense counterpart} > \textrm{ smaller dense network}$ \cite{Frankle19LotteryTicket}.}.

\begin{figure}[ht]
\centering
\includegraphics[clip,width = 0.48\textwidth]{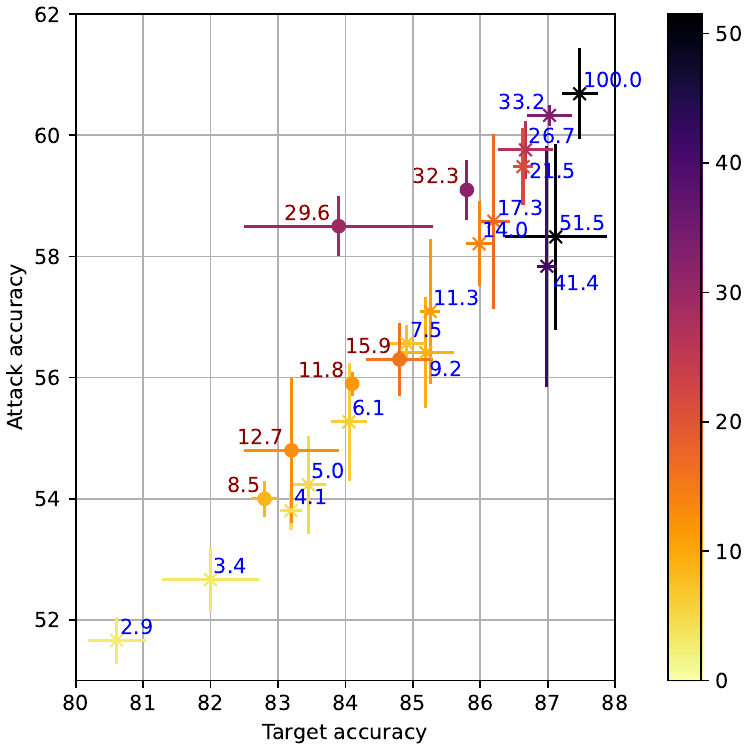}

\caption{\label{fig:phasis_diagram}
Means and standard deviations of the accuracy and defense level of various sparse networks.
The percentage of nonzero weights is given in blue for IMP ($*$~\textcolor{blue}{p$\%$}), and in red for Butterfly (\textbullet~\textcolor{red}{p$\%$}). The color indicates the percentage of nonzero parameters (in black from $50$ to $100\%$).}
\end{figure}

\paragraph{Related works.}
Whether sparsity improves privacy \emph{without further adjustment} of the training algorithm  has already been partially explored \cite{bagmar2021membership, wang2020against, yuan2022membership}. In the light of the experiments performed in this work, shortcomings of the previous approaches will be discussed in section \ref{sec:experiments}.

\begin{figure*}[t]
\centering
\includegraphics[width=0.8\linewidth]{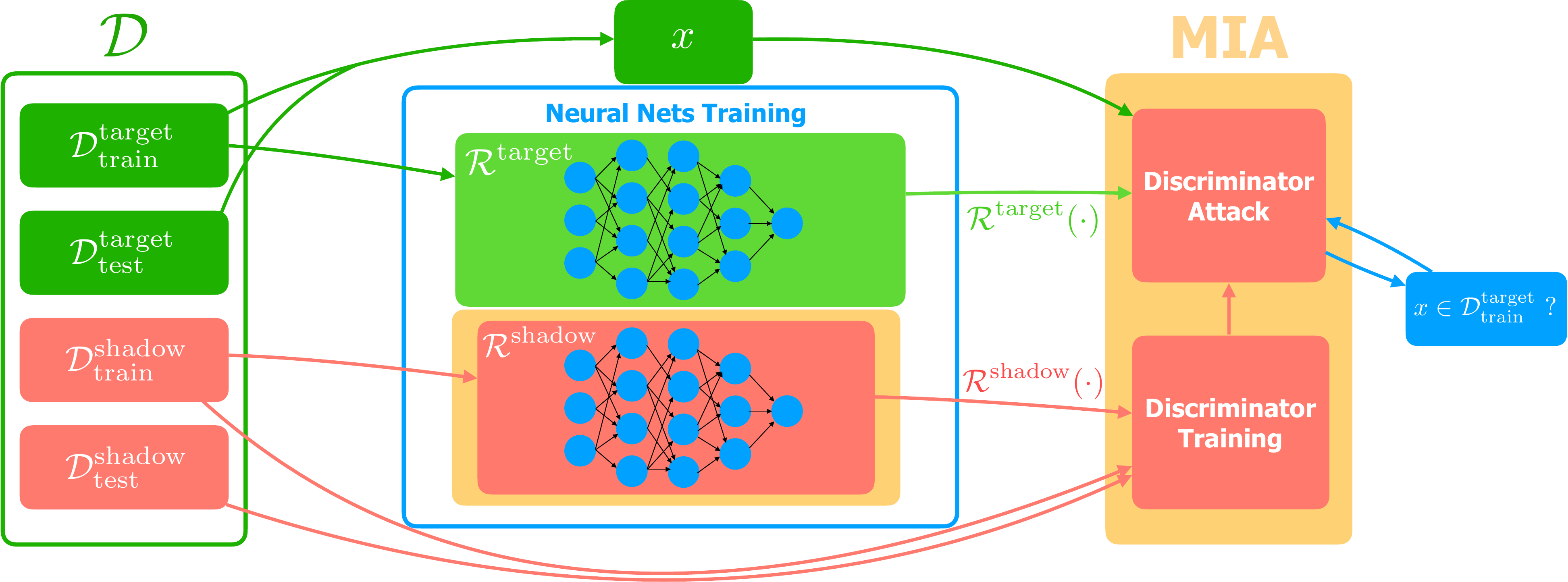}

\caption{\label{fig:mia}Experiments obey to the following pipeline: two networks are trained in the same fashion on $\Dentr^{\text{target}}$ and $\Dentr^{\text{shadow}}$ respectively. $\Rfant$, $\Dentr^{\text{shadow}}$ and $\mathcal{D}_{\text{test}}^{\text{shadow}}$ are then used to train a discriminator that will attack $\Rcible$ by trying to infer the membership of $x$ in $\Dentr^{\text{target}}$.}

\end{figure*}

\paragraph{Contributions and results.}

This work studies two types of sparsity detailed in section \ref{sec:defences}: one without any constraints on the support (location of the nonzero weights), referred to as \emph{unstructured} sparsity \cite{Frankle19LotteryTicket}, the other one being \emph{structured} sparsity \cite{dao2022monarch, DeBut}. The results in section \ref{sec:experiments} show positive correlations between the classification error of the model, its sparsity, and its privacy (measured by the attack's error, see Figure \ref{fig:phasis_diagram}).
\begin{enumerate}
    \item This is the case for unstructured sparsity, already explored in \cite{yuan2022membership}, but with different conclusions. This is discussed in section \ref{sec:experiments}.
    \item For structured sparsity, the experiments suggest a tradeoff between privacy and performance similar to the unstructured case. This is remarkable as the structure is fixed beforehand, independently of the data.
\end{enumerate}

As the presence of a correlation between the classification error of the model and its privacy is observed in the experiments, future works should design experiments that control the impact of the model's classification error in order to conclude anything on the role of sparsity. Indeed, it could be that the sparse networks encountered in the experiments are more private than their dense counterparts just because they are less accurate.

A look at previous works on sparsity and privacy shows that this fact was not taken carefully into account, see section \ref{sec:experiments}.
\paragraph{Plan} Section \ref{sec:attacks} introduces the MIAs used for the experiments. Section \ref{sec:defences} describes the types of sparsity used to defend against MIAs. Section \ref{sec:experiments} gives the results of the experiments.

\section{MIA with a shadow model}
\label{sec:attacks}
Given a dataset $\mD^{\text{target}}$, and a target network $\Rcible$ trained on a subset $\mD_{\text{train}}^{\text{target}}$ of $\mD^{\text{target}}$, a \emph{membership inference attack} (MIA) consists in infering the membership function
\[
m_{\text{target}}:x\in\mD^{\text{target}} \mapsto \left\{\begin{array}{cc}
        1 & \textrm{if } x\in \mD_{\text{train}}^{\text{target}},\\
        0 & \textrm{otherwise}.
    \end{array}\right.
\]
with only a \emph{black-box} access to the function $x\mapsto\Rcible(x)$. Most of the existing attacks try to measure the confidence of the model in its predictions made locally around $x$. If the measured confidence is high enough, then the attacker answers positively to the membership question.

In practice, the most efficient attacks consist in training a discriminator model that makes a decision based on local information of $\Rcible$ around $x$. This discriminator is trained from one or several \emph{shadow} network(s) \cite{MIAsurvey2022}, as explained below (see also Figure \ref{fig:mia}). Experiments of section \ref{sec:experiments} consider the less expensive case where the attacker trains a single shadow network.

\paragraph{Shadow network} Assume that the attacker has access to a dataset $\mD^{\text{shadow}}$ from the same distribution as $\mD^{\text{target}}$. The attacker trains their own shadow network $\Rfant$ on a subset $\mD_{\text{train}}^{\text{shadow}}$ of their data. Ideally, $\Rfant$ is trained under the same conditions as $\Rcible$ (same architecture and same optimization algorithm). At this point, the attacker has a tuple $(\Rfant, \mD^{\text{shadow}}, \Dentr^{\text{shadow}})$ which is similar to $(\Rcible, \mD^{\text{target}}, \Dentr^{\text{target}})$, and knows the shadow membership function $m_{\text{shadow}}$.

\paragraph{Discriminator} Afterwards, the attacker trains a discriminator to approximate $m_{\text{shadow}}$, given a black box access to $\Rfant$. This discriminator is then used as an approximation of $m_{\text{target}}$ given a black box access to $\Rcible$. The model for the discriminator can be any classical classifier (logistic regression, neural network, etc.) \cite{MIAsurvey2022}.

\section{Defense and neural network pruning}

\label{sec:defences}
Training sparse neural networks is first motivated by needs for frugality in resources (memory, inference time, training time, etc.).

This work investigates another property of sparsity: whether it can also improve the privacy of the training data. Note that a perfectly confidential network has not learned anything from the data and has no practical interest. Thus, a trade-off between confidentiality and accuracy must be made according to the task at hand. In what follows, two different types of sparsity are considered.

\subsection{Unstructured sparsity via IMP}

In the first case, no specific structure is imposed on the set of nonzero weights. The weights that are set to zero (pruned) are selected by an {\em iterative magnitude pruning process} (IMP) \cite{Frankle19LotteryTicket}:
(i) train a network the usual way,
(ii) prune $p\%$ of the weights having the smallest magnitude,
(iii) adjust the remaining weights by re-training the network (weights that have been pruned are masked and are no longer updated), then go back to (ii) until the desired level of sparsity is reached.

This procedure finds sparse networks with empirically interesting statistical properties \cite{Frankle19LotteryTicket, Frankle21WhyMissingTheMark}.

\subsection{Structured butterfly sparsity}
\label{sec:butterfly}

\begin{figure}[h!]
    \centering
     \begin{subfigure}[b]{0.24\linewidth}
         \centering
         \includegraphics[width=\linewidth]{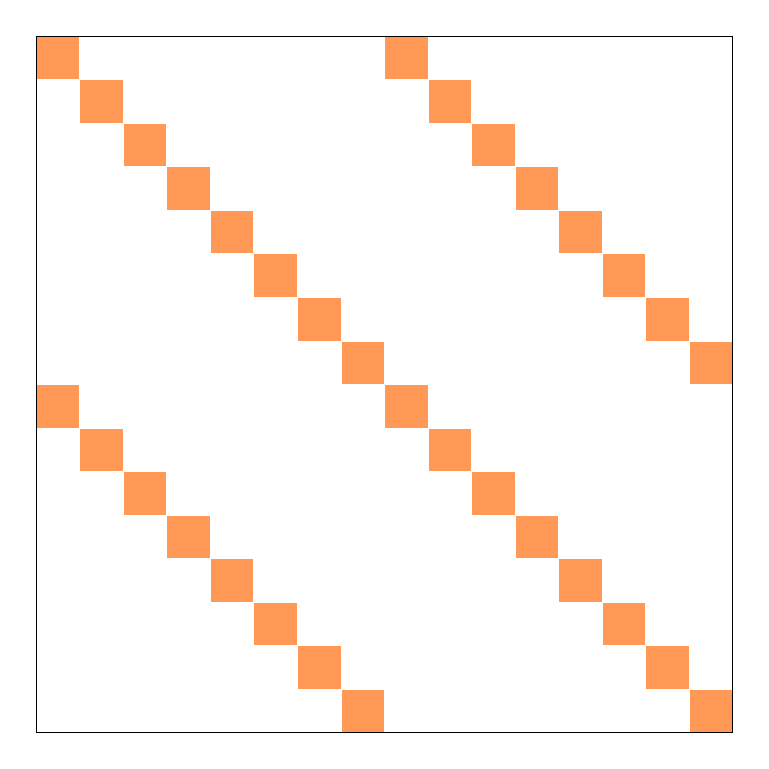}
     \end{subfigure}
     \hfill
     \begin{subfigure}[b]{0.24\linewidth}
         \centering
         \includegraphics[width=\linewidth]{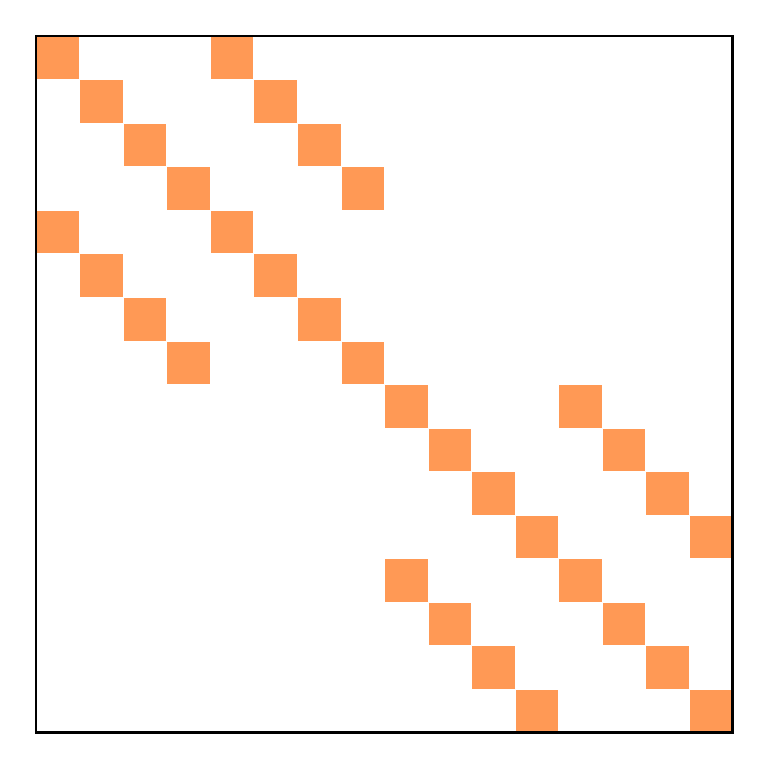}
     \end{subfigure}
      \begin{subfigure}[b]{0.24\linewidth}
         \centering
         \includegraphics[width=\linewidth]{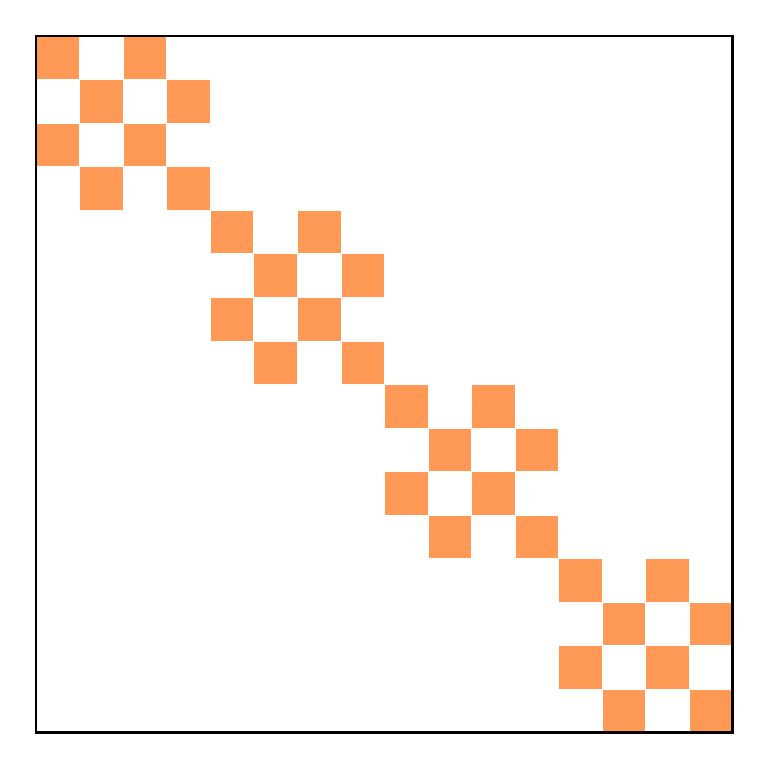}
     \end{subfigure}
      \begin{subfigure}[b]{0.24\linewidth}
         \centering
         \includegraphics[width=\linewidth]{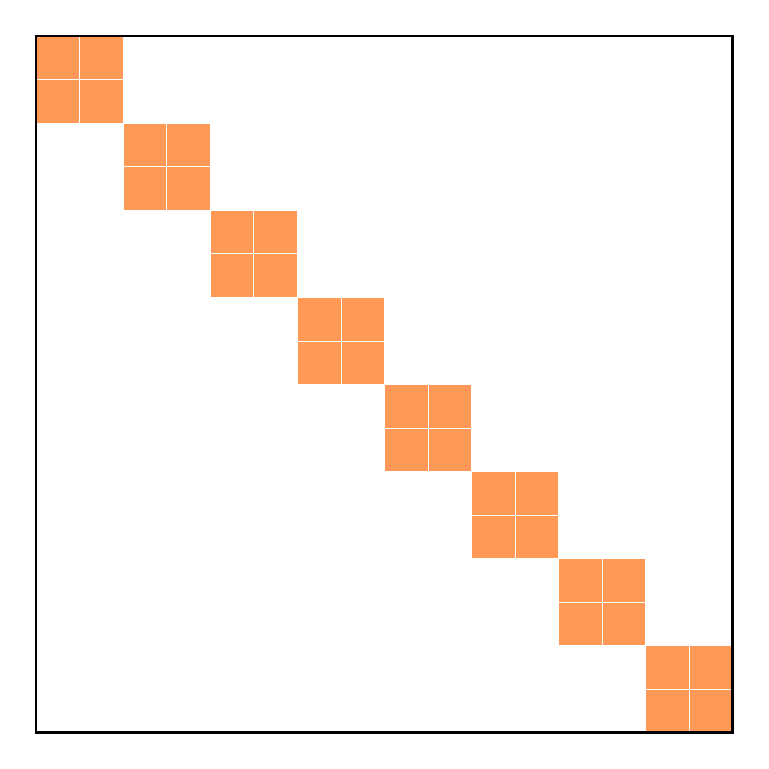}
     \end{subfigure}
     \caption{Example of supports enforced to the sparse factors in a butterfly decomposition.}
    \label{fig:butterfly}
\end{figure}

In the second case, the so-called "butterfly" sparsity is structured: the weight matrices of the neural network are constrained to be a product of sparse matrices with specific supports \cite{dao2022monarch, le2022fast, DeBut}, see Figure \ref{fig:butterfly}. This type of sparsity is particularly interesting for resource efficiency. Indeed, any $N\times N$ matrix with such a factorization has a matrix-vector multiplication complexity which is sub-quadratic in theory. It is for example $\mathcal{O}(N \log N)$ for some supports \cite{dao2019learning}, against $\mathcal{O}(N^2)$ in general.

To enforce the butterfly structure in a neural network, the weight matrices $\mathbf{W}$ are parameterized as $\mathbf{W} = \matseq{X}{1} \ldots \matseq{X}{L}$, and only the nonzero coefficients of $\matseq{X}{1}, \ldots, \matseq{X}{L}$ are initialized and then optimized by stochastic gradient descent. In the case of a convolution layer, the matrix $\mathbf{W}$ for which we impose such a structure corresponds to the concatenation of the convolutional kernels \cite{DeBut}. Butterfly networks can reach empirical performances comparable to a dense network on image classification tasks \cite{dao2022monarch, DeBut}.
\section{Experimental results}
\label{sec:experiments}

All hyperparameters (including the discriminator architecture) have been chosen via grid search, with results averaged on three experiments.

\paragraph{Dataset.}

Experiments are performed on CIFAR-10. The dataset is randomly (uniformly) partitioned into 4 subsets $\Dentr^{\text{target}}, \Dtest^{\text{target}}, \Dentr^{\text{shadow}}, \Dtest^{\text{shadow}}$ of 15000 images, respectively used to train and test the target and shadow networks. The membership functions are defined as in section \ref{sec:attacks}, with $\mD^{\text{target}}:=\Dentr^{\text{target}} \cup \Dtest^{\text{target}}$ and $\mD^{\text{shadow}}:=\Dentr^{\text{shadow}} \cup \Dtest^{\text{shadow}}$. For the target and shadow network, among their 15000 training data points, 1000 are randomly chosen and fixed for all our experiments as a validation set (used to tune the hyper-parameters, and for the stopping criterion).

\paragraph{Training of the target and shadow models.}

The target and shadow networks have a ResNet-20 architecture \cite{he2016} (272474 parameters). They are trained to minimize the cross-entropy loss by stochastic gradient descent (momentum $0.9$, no Nesterov acceleration) on their respective training sets for 300 epochs, with a batch size of 256. The dataset is augmented with random horizontal flipping and random cropping. The initial learning rate is divided by 10 after 150 and 225 epochs. The weights of the neural networks are initialized with the standard method on Pytorch, following a uniform distribution on $(-1/\sqrt{n},1/\sqrt{n})$ where $n$ is the input dimension for a linear layer, and $n$ is $\textrm{input dimension}\times \textrm{kernel width} \times \textrm{kernel height}$ for a convolution.

The initial learning rate and the weight decay are reported in table \ref{tab:hyperparameters}. These hyperparameters lead to the same performance as in \cite{he2016} when using the whole 50000 training images of CIFAR-10 instead of 15000 of them as it is done for the target and shadow networks.

For IMP, 24 prunings and readjustments of the parameters are performed. Each readjustment is done with the same training procedure as above (300 epochs, etc.). Before pruning, the weights are rewound to the values they had at the end of the epoch of maximum validation accuracy in the last $300$ epochs.

For training ResNet-20 with the butterfly structure, the original weight matrices of some convolution layers are substituted by matrices admitting a butterfly factorization, with a number $L=2$ or $3$ of factors, following a monotonic chain minimizing the number of parameters in the factorization \cite{DeBut}. The substituted layers are those of the $S=1, 2$ or $3$ last segments\footnote{A segment is three consecutive basic blocks with the same number of filters. A basic block is two convolutional layers surrounded by a residual connection.} of ResNet-20.

\begin{table}[t]
\caption{\label{tab:hyperparameters}Hyperparameters for the training of the target and the shadow neural networks.}

\begin{center}
    \resizebox{0.95\linewidth}{!}{
    \begin{tabular}{|c|c|c|c|} \hline
        Network & $\%$ of nonzero params & Initial learning rate
        & Weight decay  \\ \hline
        \multicolumn{1}{|c|}{ResNet-20 dense} & 100 \% & $0.03$ & $0.005$\\
        \multicolumn{1}{|c|}{Butterfly ($S=1, L=2$)} & 32.3 \% & 0.3 & 0.0005  \\
        \multicolumn{1}{|c|}{Butterfly ($S=1, L=3$)} & 29.6 \%  & 0.3 & 0.0001  \\
        \multicolumn{1}{|c|}{Butterfly ($S=2, L=2$)} & 15.9 \%  & 0.3 & 0.0005  \\
        \multicolumn{1}{|c|}{Butterfly ($S=2, L=3$)} & 12.9 \%  & 0.1 & 0.001  \\
        \multicolumn{1}{|c|}{Butterfly ($S=3, L=2$)} & 11.8 \%  & 0.3 & 0.0005  \\
        \multicolumn{1}{|c|}{Butterfly ($S=3, L=3$)} &  8.5 \%  & 0.1 & 0.001  \\
        \multicolumn{1}{|c|}{IMP with $k$ prunings} & $\simeq 100\times (0.8)^k\%$ & $0.03$ & $0.005$  \\ \hline
    \end{tabular}
    }
    \end{center}

\end{table}
\paragraph{Discriminator training.} A discriminator takes as inputs the class $i$ of $x$, the prediction $\Reseau(x)$ made by a network $\Reseau$ (target or shadow), as well as an empirical approximation of $ \frac{1}{\epsilon} \mathbb{E} \left(|\Reseau(x) - \Reseau(x + \epsilon \mathcal{N}) | \right)$ for $\epsilon = 0.001$ and a standard normal vector $\mathcal{N}$, where the approximation is made by averaging over 5 samples. This expectation encodes local first order information of $\Reseau$ around $x$. For each pair of networks ($\Rcible$,$\Rfant$), three discriminators (perceptrons) are trained, with respectively 1, 2, 3 hidden layer(s) and 30, 30, 100 neurons on each hidden layer. The binary cross entropy is minimized with Adam for 80 epochs, without weight decay and for three different learning rates $\{0.01, 0.001, 0.0001\}$.

\paragraph{Privacy and performance metrics} The performance of the network is measured by its accuracy (percentage of the data correctly classified), and the privacy is measured by the accuracy of the attacker. In our case, there are as much seen as unseen data during training of the network (target or shadow). Ideally, the discriminator should not do better than guessing randomly, having then an accuracy of $50\%$. When the discriminator has a higher accuracy, this means that the training data is less confidential.

\paragraph{Results}

The experiments show that the privacy improves when both the sparsity goes up and the performance goes down. This shows that experiments of this type cannot be sufficient to conclude on the role of sparsity for privacy, since the performance is clearly correlated with the privacy, and it could be the main explicating variable. This suggests that further experiments are needed, for instance by keeping the performance of the sparse networks at a constant level. Moreover, larger scale experiments seem to be needed to reduce the standard deviations observed on the privacy metric (the attack accuracy).

\paragraph{Caveats of previous works.} In the light of Figure \ref{fig:phasis_diagram}, the experiments that assess the role of sparsity for privacy should take into consideration:
\begin{enumerate}
    \item [(1)] the possible presence of a correlation between the accuracy of the model and the privacy;
    \item [(2)] the possibly high standard deviations of the privacy metric.
\end{enumerate}
To the best of our knowledge, the three papers \cite{bagmar2021membership, wang2020against,yuan2022membership} are the only works that investigate the role of sparsity for privacy. Unfortunately, points (1) and (2) are not taken into account in previous approaches, as discussed below.

The experimental results in \cite{yuan2022membership} suggest that increasing the sparsity \emph{degrades} privacy. However, the Figure 7 of \cite{yuan2022membership} is subject to point (1) for most of the reported configurations, and  subject to point (2) in all configurations. Indeed, it shows that most of the sparse networks that are less private than their dense counterparts are also more accurate. This is exactly point (1), and it is thus impossible to rule out that it is the increased accuracy, rather than the increased sparsity, that may cause the loss in privacy. For the few models of Figure 7 in \cite{yuan2022membership} that do not suffer from point (1), they still suffer from point (2). Standard deviations are not reported in \cite{yuan2022membership}, while Figure \ref{fig:phasis_diagram} show high standard deviations in a similar setup.

In the results of \cite{wang2020against}, no obvious correlations can be inferred between the accuracy of the model and its privacy. But the absence of standard deviations makes any conclusion on the role of sparsity hard to draw.

The results of \cite{bagmar2021membership} suggest that lottery ticket networks have identical privacy risks as their dense counterparts. Yet the classification accuracy of the networks was not reported and not taken into account. Moreover no standard deviations are displayed in the figures presenting the results. Again, it is hard to conclude from these experiments.

\section{Conclusion}

Sparsity is firstly motivated by resource efficiency. This work investigates whether it can also be used for privacy concerns in the case of unstructured sparsity (IMP) and structured sparsity (butterfly). The experiments show a positive correlation between the classification error of the model, its sparsity and its privacy. These results led to the identification of two important points that should be taken into account in future works that investigate sparsity and privacy: 1) the potential presence of a correlation with the classification error and 2) the potentially high variability in the privacy metric. Moreover, attacks that are more statistically relevant, but also more expensive, should be considered in the future \cite{Carlini22LIRA}.


{\scriptsize
\begin{spacing}{0.8}

\end{spacing}
}



\begin{thebibliography}{10}

\bibitem{bagmar2021membership}
Aadesh~Mahavir Bagmar, Shishira Maiya, Shruti Bidwalkar, and Amol Deshpande.
\newblock Membership inference attacks on lottery ticket networks.
\newblock In {\em ICML 2021 Workshop on Adversarial Machine Learning}, 2021.

\bibitem{Belkin19DoubleDescent}
M.~Belkin, D.~Hsu, S.~Ma, and S.~Mandal.
\newblock Reconciling modern machine-learning practice and the classical
  bias-variance trade-off.
\newblock {\em National Academy of Sciences USA}, 2019.

\bibitem{Carlini22LIRA}
Nicholas Carlini, Steve Chien, Milad Nasr, Shuang Song, Andreas Terzis, and
  Florian Tram{\`{e}}r.
\newblock Membership inference attacks from first principles.
\newblock In {\em 43rd {IEEE} Symposium on Security and Privacy, {SP} 2022, San
  Francisco, CA, USA, May 22-26, 2022}, pages 1897--1914. {IEEE}, 2022.

\bibitem{dao2022monarch}
T.~Dao, B.~Chen, N.~S. Sohoni, A.~Desai, M.~Poli, J.~Grogan, A.~Liu, A.~Rao,
  A.~Rudra, and C.~R{\'e}.
\newblock Monarch: Expressive structured matrices for efficient and accurate
  training.
\newblock In {\em ICML}, 2022.

\bibitem{dao2019learning}
T.~Dao, A.~Gu, M.~Eichhorn, A.~Rudra, and C.~R{\'e}.
\newblock Learning fast algorithms for linear transforms using butterfly
  factorizations.
\newblock In {\em ICML}, 2019.

\bibitem{Frankle19LotteryTicket}
J.~Frankle and M.~Carbin.
\newblock The lottery ticket hypothesis: Finding sparse, trainable neural
  networks.
\newblock In {\em ICLR}, 2019.

\bibitem{Frankle21WhyMissingTheMark}
J.~Frankle, G.~K. Dziugaite, D.~Roy, and M.~Carbin.
\newblock Pruning neural networks at initialization: Why are we missing the
  mark?
\newblock In {\em ICLR}, 2021.

\bibitem{he2016}
K.~He, X.~Zhang, Sh. Ren, and J.~Sun.
\newblock Deep residual learning for image recognition.
\newblock In {\em CVPR}. IEEE, 2016.

\bibitem{MIAsurvey2022}
H.~Hu, Z.~Salcic, L.~Sun, G.~Dobbie, P.~S. Yu, and X.~Zhang.
\newblock Membership inference attacks on machine learning: A survey.
\newblock {\em ACM Computing Surveys}, 2022.

\bibitem{le2022fast}
Q.-T. Le, L.~Zheng, E.~Riccietti, and R.~Gribonval.
\newblock Fast learning of fast transforms, with guarantees.
\newblock In {\em ICASSP}. IEEE, 2022.

\bibitem{DeBut}
R.~Lin, J.~Ran, K.~H. Chiu, G.~Chesi, and N.~Wong.
\newblock Deformable butterfly: A highly structured and sparse linear
  transform.
\newblock In {\em NeurIPS}, 2021.

\bibitem{quemener2013}
E.~Quemener and M.~Corvellec.
\newblock {SIDUS}—the {Solution} for {Extreme} {Deduplication} of an
  {Operating} {System}.
\newblock {\em Linux Journal}, 2013.

\bibitem{SurveyRigaki20}
Maria Rigaki and Sebastian Garcia.
\newblock A survey of privacy attacks in machine learning.
\newblock {\em CoRR}, abs/2007.07646, 2020.

\bibitem{shokri2017membership}
R.~Shokri, M.~Stronati, C.~Song, and V.~Shmatikov.
\newblock Membership inference attacks against machine learning models.
\newblock In {\em SP}. IEEE, 2017.

\bibitem{tan2023blessing}
J.~Tan, D.~LeJeune, B.~Mason, H.~Javadi, and R.~G. Baraniuk.
\newblock A blessing of dimensionality in membership inference through
  regularization.
\newblock In {\em AISTATS}, 2023.

\bibitem{wang2020against}
Y.~Wang, C.~Wang, Z.~Wang, S.~Zhou, H.~Liu, J.~Bi, C.~Ding, and S.~Rajasekaran.
\newblock Against membership inference attack: Pruning is all you need.
\newblock {\em IJCAI}, 2021.

\bibitem{yuan2022membership}
X.~Yuan and L.~Zhang.
\newblock Membership inference attacks and defenses in neural network pruning.
\newblock In {\em USENIX}. IEEE, 2022.

\bibitem{Zhang21UnderstandingRequiresRethinking}
C.~Zhang, S.~Bengio, M.~Hardt, B.~Recht, and O.~Vinyals.
\newblock Understanding deep learning (still) requires rethinking
  generalization.
\newblock {\em ACM}, 2021.

\end{thebibliography}
\end{document}